%% file: 5221RS.tex
\font\caps=cmcsc10 at 12pt
\font\eightrm=cmr8 at 8pt
\documentclass{oxarticle}
\usepackage{amsmath, amssymb}

\input{Johnmacros22}
\newcounter{orange} 
\newcounter{apple} 
\addtocounter{apple}{1}

\newcounter{grape} 
\addtocounter{grape}{1}
\newcommand{\numberhere}{5221}
\newcommand{\articlenumber}{\numberhere{RS}}

\proofmodefalse
\usepackage{color} 

\begin{document}

 \begin{center}

{ \Huge Suppressed SUSY for the SU(5) Grand Unified Supergravity Theory\\[.5cm] }  

\vspace*{.1in}
%

\renewcommand{\thefootnote}{\fnsymbol{footnote}}

{\caps John A. Dixon\footnote{jadixg@gmail.com}\\CAP\\Edmonton, Canada 
} \\[.5cm] 
\end{center}
\Large
\normalsize 

 \begin{center}
 Abstract 
\end{center}

This paper starts with the most basic SU(5) Grand Unified Theory, coupled to Supergravity. Then it builds a new theory, incorporating the ideas of Suppressed SUSY. Suppressed SUSY is an alternative to the spontaneous breaking of SUSY.  It does not need an invisible sector or explicit soft breaking of SUSY.  It varies the content of the supermultiplets while  keeping the restrictive nature of SUSY.  For the simple model and sector constructed here, Suppressed SUSY has only three  dimensionless parameters, plus the Planck mass.  At tree level, this predicts a set of 8 different new masses, along with a cosmological constant that is naturally zero.  The X and Y vector bosons get Planck scale masses $2 \sqrt{10} g_5 M_{\rm P}$.  The five scalar multiplets that accompany the Higgs, and the Gravitino, all get colossally huge `SuperPlanck' scale masses of order $M_{\rm SP} \approx 10^{17} M_{\rm P}$ from a see-saw mechanism that arises from the theory.  This new mass spectrum, the well-known $SU(5)$  weak angle problem, and  the cosmological constant value, should serve as guides for further modifications for the new Action.

\refstepcounter{orange}
{\bf \theorange}.\;
{\bf SUSY GUTs tend to be unpredictive:}  
Grand Unified Models coupled to \SG\ tend to be unpredictive, because there are so many possibilities  \ci{freepro}.  Added to this are the problems from spontaneous SUSY breaking, with its sum rules and fine-tuning \ci{ferrarabook,superspace,WB,Buchbinder:1998qv,xerxes,west,Weinberg3,haber,buchmueller}.  These have given rise to the ideas of  the invisible sector and explicit soft breaking,
which are incorporated into the `Minimal Supersymmetric Standard Model' (`MSSM'), with over a hundred new parameters  \ci{haber,buchmueller}.

\refstepcounter{orange}
{\bf \theorange}.\;
{\bf The MSSM  has not yet been confirmed experimentally:}  
Even using many parameters, attempts to understand the elementary particles or astrophysics using the MSSM  have, so far, not met with success.  Many recent experiments testing the MSSM at the LHC have been fruitless \ci{buchmueller}.
There is a mounting belief that Squarks and Gluinos, for example, will never be found experimentally  \ci{madrid}.

\refstepcounter{orange}
{\bf \theorange}.\;
{\bf Two `Obvious Features' of SUSY:}  
\la{obvious}
It seems obvious that if Supersymmetry (`SUSY')  is relevant to physics, then superpartners should exist and be detectable.  
It also seems obvious that there should be a mass splitting between particles and their superpartners by way of spontaneous breaking of SUSY. These obvious conclusions stem from the incorporation of the SUSY algebra into the transformations of the quantized fields of the theory.

 \refstepcounter{orange}
{\bf \theorange}.\;
{\bf  Suppressed SUSY provides a way to avoid these `Obvious Features':} 
Like every other gauge theory,  local SUSY gives rise to a Master Equation \ci{poissonbrak,Becchi:1975nq,zinnbook,taylor,Zinnarticle,becchiarticles1,becchiarticles2,BV,Weinberg2}. The Master Equation incorporates the intricate, complicated and highly restrictive nature of local SUSY.  This paper shows, using an example, how we can keep the Master Equation, but change the quantum field theory, by using an `Exchange Transformation' \ci{four,five,goldstein, LandauLifmechanics,nonrermorm}.  An \ET\ has the form of a canonical transformation of the Master Equation, but it interchanges sources and fields.  The result is a different theory with a new Master Equation, because fields are quantized and sources are not.  The problems  \ci{pran,moultaka,ferrarasagnoti,Weinberg3} arising from the need for the spontaneous breaking of SUSY are avoided.  But the intricate, complicated and highly restrictive nature of local SUSY survives, and it governs the physics in  detail.

\refstepcounter{orange}
{\bf \theorange}.\;
{\bf Example of an \ET:}  
The \ET\ used here is set down in Paragraphs \ref{genfunc} and \ref{genfunc2} below.  The \ET s are always linear, homogeneous, and invertible. So the new Action and Master Equation are `formally identical' to the old ones, except that the symbols and properties for some of the fields and sources get changed. 
As a result, the new Action is divided into a quantized Action and a source Action in a different way from the original Action.

 \refstepcounter{orange}
{\bf \theorange}.\;
{\bf The Scalar Potential for \SG:}  
For the simplest case\footnote{Coupling \SG\ to chiral matter is a surprisingly  complicated subject, as the authors of \ci{freepro} candidly acknowledge.  A little confidence, and a check on normalizations, can be gained by applying this formula (\ref{complexV1}), and also its variation in  (\ref{complexV2}) and (\ref{defnN}), to get formula (19.27) of \ci{freepro}, as discussed on page 398 of \ci{freepro} for the Polonyi model.}, the scalar potential function $V$ for supergravity has the form \ci{freepro}:
    \be
 V = e^{\lt (\k^2 \sum_i z^i \ov z_i\rt )}
 \lt \{ \sum_j \lt | \fr{\pa W}{\pa z^j}+ \k^2 {\ov z}_j W \rt |^2
- 3 \k^2 \lt | W \rt |^2 \rt \}
+ \fr{1}{2} \sum_{\a} \lt | D_{\a}\rt |^2
\la{complexV1}  
\ee
	Here $W(z)$ is the superpotential,  $D_{\a}(z, \ov z)$ are the auxiliary $D$ terms for the Yang-Mills theory, and $z^i$ are the chiral scalar fields in the action, with complex conjugates $\ov z_i$. The constant $ \fr{1}{\k} = \fr{M_{\rm Planck} }{\sqrt{8 \pi}}= 2.4 \times 10^{18}\; {\rm GeV}= M_P$  is the `reduced' Planck mass\footnote{Quoting p 574 of \ci{freepro}, we note that $\k^2 = \sqrt{8 \pi G}$ and $M_{\rm Planck} = G^{-\fr{1}{2}}= 1.2 \times 10^{19}$  GeV. }.

 \refstepcounter{orange}
{\bf \theorange}.\;
{\bf V and the \cc:}  
This  $V$ in (\ref{complexV1}) is the scalar non-derivative term in  $-e^{-1} L$ where $L$ is the Lagrangian and e is the vierbein determinant\footnote{The universal but confusing practice is that the letter $e$  stands for $e =2.71828$ as well as the determinant of the vierbein, depending on context.}.  Clearly, if the Vacuum Expectation Value (`VEV'), $\lt < V \rt > \neq 0$, this term yields a `cosmological constant term'  $\int d^4 x\; e \lt <- V \rt >$ in the
 action. 

\refstepcounter{orange}
{\bf \theorange}.\;
{\bf V is not positive definite:}  
The $V$ in equation (\ref{complexV1}) is manifestly not positive definite because of the term $- 3 \k^2 \lt | W \rt |^2 $. 
The discovery that the scalar potential (\ref{complexV1})  is not positive definite quickly led to a number of efforts to find a spontaneously broken theory of SUSY with zero cosmological constant.  
The negative term 
$\lt < - 3 e^{\lt (\k^2 \sum_i z^i \ov z_i\rt )} \k^2 \lt | W \rt |^2 \rt >$ in (\ref{complexV1}) could be fine tuned against the  positive term from the non-zero auxiliary VEV contributions in (\ref{complexV1}), so that the net contribution to $\lt < V \rt >$ was zero.  Since it is known that the cosmological constant is very tiny, although perhaps non-zero \ci{ccnonzero,darkenergy}, this was deemed necessary to make contact with cosmology.

\refstepcounter{orange}
{\bf \theorange}.\;
{\bf Development of the New  `Counting Form' Theory with a naturally zero cosmological constant:}  
 This paper constructs a theory with Suppressed SUSY, starting from the \GUST.  The crucial step is  the rewriting of $V$ in the  `Counting Form'  (\ref{complexV2}) below.  This suggests 
  the special superpotential in Paragraph \ref{specialsuperpotential} and the   \ET\ used in Paragraph \ref{summaryETs}, and explained more fully in 
  Paragraphs \ref{genfunc} and \ref{genfunc2}. Detailed Mathematica calculations of the mass spectrum are described, starting in Paragraph \ref{mathematicaremarks}.  Some general remarks about the Master Equation start in Paragraph \ref{mastereqpara}, and then the detailed results are  summarized in  Paragraph \ref{conclusion}.

\refstepcounter{orange}
{\bf \theorange}.\;
{\bf Alternate Form of the Scalar Potential $V$}  
The expression (\ref{complexV1}) can also be written:
\be
V
=
e^{\lt (\k^2 \sum_i z^i \ov z_i\rt )} \lt \{
 \sum_j \lt | \fr{\pa W}{\pa z^j} \rt |^2
 + 
 \k^2  
  \lt [
 N    
- 3     
+\k^2 
\sum_j z^j
{\ov z}_j 
 \rt ]
\lt | W 
 \rt |^2
\rt \}
+ \fr{1}{2} \sum_{\a} \lt ( D_{\a}\rt )^2
\la{complexV2}
\ee
where we define the Field Counting Operator:
\be
N=
 \sum_j 
 \lt (
  z^j  \fr{ \pa}{\pa z^j}
 +
 \ov z_j  \fr{\pa}{\pa \ov z_j}  
\rt ) 
\la{defnN}
\ee

\refstepcounter{orange}
{\bf \theorange}.\;
{\bf Just the Higgs Sector:}  
\la{justhiggs}
For the Higgs sector of the model of interest here the foregoing can be written \ci{ross,GUT}:
\be
V
=
e^{\k^2 \lt ( \sum_{i=1}^5 \lt (H^i_L \oH_{Li}+ H_{Ri} \oH_R^i \rt )
+ \sum_{i=1}^{24} S^a \oS^a  \rt )} 
\lt \{
 \sum_{i=1}^5 \lt ( 
 \lt | \fr{\pa W}{\pa H^i_L }\rt |^2  
 + \lt |  \fr{\pa W}{\pa H_{Ri} }\rt |^2  
 \rt )
 +
  \sum_{i=1}^{24}  
 \lt |  \fr{\pa W}{\pa S^a }\rt |^2  
\la{firstlineofVfromHbeforeET}
\ebp
 + 
 \k^2  
  \lt [
 N    
- 3     
+\k^2 
\lt ( \sum_{i=1}^5 \lt (H^i_L \oH_{Li}+ H_{Ri} \oH_R^i \rt )
+ \sum_{a=1}^{24} S^a \oS^a  \rt ) \rt ]
\lt | W 
 \rt |^2
\rt \}
+  \fr{1}{2} \sum_{a=1}^{24}\lt (D^a\rt )^2
\la{complexV2here}
\ee
where $D^a$ can be found below in Equation (\ref{Diszero}), and
\be
N=
  \sum_{i=1}^5 \lt (H^i_L \fr{\pa}{\pa H_{L}^i }
 + H_{Ri} \fr{ \pa}{\pa H_R^i } \rt )
+ \sum_{i=1}^{24} \lt ( S^a \fr{\pa}{\pa S^a}  \rt ) 
 +*
\la{counthiggs}\ee

 The fields here are the scalar terms for the  usual chiral Higgs type superfields. 
   The scalar field  $S^a$ is in the adjoint of SU(5),   $H_L^i$  
   is a  5 of SU(5), and $H_{Ri}$ is a ${\ov 5}$ of SU(5).  To be scalars in a chiral superfield, all of these need to be complex, including the  $S^a$ (even though the adjoint is a real representation).   Note:  
\be
S^{\;i}_{j}= S^a T^{ai}_j; \; T^{ai}_j T^{bj}_i = 2 \d^{ab} ; \;   \lt [T^{a}, T^{b}\rt]   = i f^{abc}T^{c}    
;\;\Tr[S^2] = 2 S^a S^a;\; T^a = (T^a)^{\dag}
 \ee

\refstepcounter{orange}
{\bf \theorange}.\;
{\bf A Special Form for the Superpotential:} 
\la{specialsuperpotential} 
Here is the superpotential $W$ that we will use here for the Higgs sector. This is designed so that the \cc\ will be zero after the \ET s and gauge symmetry breaking:  
\be
 W =
 e^{\fr{-1}{4}\k^2 \lt (2 H_{L}^i H_{Ri} + \fr{1}{2}\Tr S^2 
  \rt )} 
M_P^{\fr{3}{2}}  \sqrt{g_1 \Tr( S S S) - g_2 H_L^i S_i^{\;j}  H_{Rj}  }   
\la{fresefas}
    \ee

  The factor   $M_P^{\fr{3}{2}}$ is needed to get the dimensions right.  The parameters $g_1$ and $g_2$ are dimensionless. 
   This $W$ is clearly quite strange, but it satisfies some strange requirements, explained in the rest of the paper\footnote{We would add 
   \be
 W_{\rm Matter} =
 g_3 H_L^i  T_{ij}   F^j
 + g_4 H_{Ri}  T_{jk} T_{lm} \ve^{ijklm}   + \k g_6 H_L^i S_i^l T_{lj}   F^j
 + \k g_7 H_{Rl}S_i^l  T_{jk} T_{lm} \ve^{ijklm}   + O(\k^2)
 \ee
 to include the matter here. As usual in SU(5) theories, the Squarks and Sleptons are in F and T, 
  where  $F^i$  is another 5, and $T_{ij}$ is the antisymmetric $\ov{10}$. The dimensionless constants   $g_3 ,g_4, g_6 ,g_7 $  are really matrices in the flavour space of the quarks and leptons, and we will discuss them very little here. We would need to add suitable terms in the exponential in (\ref{complexV2}) also.   Anyway, the \ET s will remove the scalars $F^i$ and $T_{ij}$  from the theory, 	but the related fermion terms do remain of course.  The matter scalars get transformed to Zinn sources, since they are Squarks and Sleptons. Similar techniques were used in  \ci{four,five}. But those papers were not based on \SG, and there is a big difference, as we see here.}.
 There are really just two masses here which we could write under the square root as $M_1^3=g_1 M_P^3$ and $M_2^3=g_2 M_P^3$. 
 
\refstepcounter{orange}
{\bf \theorange}.\;
{\bf
Action of the Counting Operator $N$:}
    The reason for choosing the square root in equation (\ref{fresefas})  is that the Counting Operator 
    (\ref{counthiggs})
    works simply with the resulting dimensions in $W$:
   \be
  \lt (
 N    
- 3    \rt )  
\lt |\sqrt{g_1 \Tr( S S S) - g_2 H_L^i S_i^{\;j}  H_{Rj}  } \rt |^2 =0
\la{nminus3}
\ee  

This follows because $N$ counts fields\footnote{ \la{ftnt1} More exotic solutions exist.  For example $  \lt (
 N    
- 3    \rt )  \lt | \sqrt{\fr{{\rm Tr} (S^5)}{H_L \cdot H_R}} \rt |^2=0$.  
But it seems that this will have undesirable singularities  in field space when $H_L \cdot H_R \ra 0$.   However  $  \lt (
 N    
- 3    \rt )  \lt | \sqrt{\fr{{\rm Tr} (S^5)}{ {\rm Tr}  (S^2):}} \rt |^2=0$   
is less pathological, and combinations of solutions like this should probably be considered.  Here we look at the simplest case for a start.}
, and the number of fields in the expression is $2 \times \fr{3}{2}=3$.
Next we note that for any constant $a$:
\be
\lt [ N,
e^{ a\k^2 \lt (2 H_{L}^i H_{Ri} + \fr{1}{2}\Tr S^2 \rt )} 
 \rt ]
 =
2 a\k^2 \lt (2 H_{L}^i H_{Ri} + \fr{1}{2}\Tr S^2 \rt )e^{ a\k^2 \lt (2 H_{L}^i H_{Ri} + \fr{1}{2}\Tr S^2 \rt )} 
\la{countexp}  \ee

\refstepcounter{orange}
{\bf \theorange}.\;
{\bf
An example of an \ET:}
\la{summaryETs}
To get the Suppressed SUSY field theory from the above \GUST, we will implement the field dependent part of a simple but important \ET, as follows. A more complete explanation of this can be found below in paragraph \ref{genfunc}:
\be
H_L^i \ra 
\fr{1}{\sqrt{2}}
H^i
,\; 
\oH_{Li} \ra 
\fr{1}{\sqrt{2}}
\oH_i
,\; 
H_{R i} \ra  
\fr{1}{\sqrt{2}}
 \oH_i  
,\; 
\oH_{R}^{ i} \ra  
\fr{1}{\sqrt{2}}
 H^i  
,\; S^a \ra 
\fr{1}{\sqrt{2}}
 K^a
,\; \oS^a \ra 
\fr{1}{\sqrt{2}}
 K^a
\la{fieldetsforhiggs}
\ee

We also eliminate all of the Sleptons and Squarks and Higgsinos and Gauginos:
\be
F^i \ra 0
,\; T_{ij} \ra 0
;\;
 \c_{L}^i \ra 0,     \c_{R}^i \ra 0,    
     \c_{S j }^i \ra 0,  \lam^a \ra 0  
    \ee
So the result is that the theory retains, as quantized fields,  only half of the original Higgs Bosons, plus 
the Leptons, Quarks, and Gauge Bosons, plus the Gravitino and the Graviton.
All the fields we set to zero are  taken into Zinn sources, so they are really still there, but performing a different role. The total action, including all the Zinn source terms, is really the same, but with some changes of names and roles.

\refstepcounter{orange}
{\bf \theorange}.\;
{\bf New Form of $W$:}
The above transformations on (\ref{fresefas}) yield the simpler, and real,  expression  $W_R$:

\be
W\ra W_{\rm R}  = e^{-\frac{1}{4}  \kappa ^2  \left(  \fr{1}{4} \Tr( K K) +  H \cdot \oH   \right) }  M_P^{\fr{3}{2}}  2^{-\fr{3}{4}}
\sqrt{ g_1  \Tr( KKK) - g_2 H \cdot K \cdot  \oH      }   
 \la{gensolsu5}
    \ee
The rest of the original expression is still present, but we can ignore it for now because it contains Zinn sources which do not get quantized.

\refstepcounter{orange}
{\bf \theorange}.\;
{\bf The Kahler Metric:}  
 Note that the  Kahler  metric term 
in (\ref{firstlineofVfromHbeforeET}) becomes:
\be
 \k^2  \left(  \fr{1}{2} \Tr( S \oS^T) +  H_L \cdot \oH_L +  H_R \cdot \oH_R \right )
\ra \k^2 
 \lt (H^i \oH_{i} + \fr{1}{4}\Tr K^2 \rt )
\la{fromKahler}\ee
We get the same expression from the term in the exponential in the superpotential (\ref{fresefas}):
\be
 \k^2  \lt (2 H_{L}^i H_{Ri} + \fr{1}{2}\Tr S^2 \rt ) \ra \k^2 
 \lt (H^i \oH_{i} + \fr{1}{4}\Tr K^2 \rt )
\la{expgoestorealpieces}
 \ee

\refstepcounter{orange}
{\bf \theorange}.\;
{\bf The Scalar Potential Simplifies:}  
The field dependent part of the scalar potential (\ref{complexV2here})  after this \ET\ is  of the form:  
\la{simpscalarpot}
  \be
V_{\rm R}
=
e^{ \k^2 \lt (H^i \oH_{i} + \fr{1}{4}\Tr K^2 \rt )}  
 \lt \{
\lt (
 \sum_{i=1}^5 
 \lt | \fr{\pa  W_{\rm R} }{\pa H^j} \rt |^2
+ 
\fr{1}{2} \sum_{a=1}^{24} \lt( \fr{\pa  W_{\rm R} }{\pa K^a} \rt )^2
\rt )
\la{firstlineofVr}
\ebp
+ \k^2    W_{\rm R}
\lt [
2  N_{\rm Real}    
- 3     
+\k^2 
 \lt (H^i \oH_{i} + \fr{1}{4}\Tr K^2 \rt )
 \rt ] W_{\rm R}
\rt \}
+ \fr{g_5^2}{2} \sum_{a=1}^{24}
\lt ( H^jT^{ai}_j \oH_{i}  - \oH_{i} T^{ai}_j H^{j}   + i f^{abc} K^b K^c\rt )^2
\la{complexV44}
\ee
In the above, as far as the fields are concerned, using the definitions in (\ref{genfunc}):
\be
N \ra  N_{\rm Real}    =
H^i \fr{\pa}{H^i}+ \oH_i \fr{\pa}{\oH_i} + K^a \fr{\pa}{K^a}   \ee

Then, using identities like (\ref{countexp}) and
(\ref{nminus3}), it is easy to show that
 the following is true:
\be
\lt [
2 N_{\rm Real}   
- 3     
+\k^2 
 \lt (H^i \oH_{i} + \fr{1}{4}\Tr K^2 \rt )
 \rt ]
W_R\equiv 0
\la{requirementtogetbacktorigidstuff}
\ee
 Recall (\ref{complexV2here}) and (\ref{fieldetsforhiggs}). We note that:
\be
  D^a  = g_5  \lt (H^j_L T^{a i}_j \oH_{Li}  - H_{Ri} T^{ai}_j \oH_R^{i}   
+i f^{abc}S^b \oS^c \rt )
\ra
g_5 \lt ( H^jT^{ai}_j \oH_{i}  - \oH_{i} T^{ai}_j H^{j}   + i f^{abc} K^b K^c\rt ) \equiv 0 
\la{Diszero}
\ee
because $K^a$ is real and $ f^{abc}$ is antisymmetric.  For $H^i$ the two contributions from $H_L$ and $H_R$ in  (\ref{fieldetsforhiggs}) cancel each other in (\ref{Diszero}).

\refstepcounter{orange}
{\bf \theorange}.\;
{\bf Simplification of V and Generation of VEVs:}
\la{thesimpleVR}
Given the form (\ref{gensolsu5}), (\ref{requirementtogetbacktorigidstuff}) and (\ref{Diszero}) become identically zero for the fields. 
So now  (\ref{firstlineofVr}) and (\ref{complexV44}) reduce to just the first line (\ref{firstlineofVr}). 
This form suggests that we can look for nontrivial solutions  where this potential $V_R$ is evaluated at the VEVs of the Scalar fields, after shifts like (\ref{Shifts}) below, as in (\ref{veviszero2}).

\refstepcounter{orange}
{\bf \theorange}.\;
{\bf
Mathematica Calculations:}
\la{mathematicaremarks}
Most of the calculations in this paper were done using Mathematica. A Mathematica Notebook which contains, and explains, the various calculations, is supplied in  \ci{mathnotebooks}.  There are 34 real scalar fields, and 24 real vector fields. These mix together in complicated ways, and that is a lot to keep track of. The full expressions for the various invariants are large. The calculation time is quite short (5 minutes).
	A fundamental technique, which is crucial here, is the expansion of series in the tiny dimensionless parameter  	 $f$ in Equation (\ref{valueoff}).

\refstepcounter{orange}
{\bf \theorange}.\;
{\bf
Detailed Notation  used in Mathematica:}
\la{notationdetails}
Now we convert the scalars to real component fields as follows ($K_i^{\; j} =\sum_{a=1}^{24}K_a T_i^{a j}$):
\be
K_i^{\; j} =
\left(
\begin{array}{ccccc}
 K_3+\frac{K_8}{\sqrt{3}}-
 \frac{2 K_{12}}{\sqrt{15}} & K_1+i K_2
   & K_4+i K_5 & K_{13}+i K_{14}
   & K_{19}+i K_{20} \\
 K_1-i K_2 &
   -K_3+\frac{K_8}{\sqrt{3}}-\frac{2 K_{12}}{\sqrt{15}} & K_6+i
   K_7 & K_{15}+i K_{16} &
   K_{21}+i K_{22} \\
 K_4-i K_5 & K_6-i K_7 & -\frac{2
   K_8}{\sqrt{3}}-\frac{2
   K_{12}}{\sqrt{15}} & K_{17}+i
   K_{18} & K_{23}+i K_{24} \\
 K_{13}-i K_{14} & K_{15}-i
   K_{16} & K_{17}-i K_{18} &
   K_{11}+\sqrt{\frac{3}{5}}
   K_{12} & K_9+i K_{10} \\
 K_{19}-i K_{20} & K_{21}-i
   K_{22} & K_{23}-i K_{24} &
   K_9-i K_{10} &
   \sqrt{\frac{3}{5}}
   K_{12}-K_{11} \\
\end{array}
\right)
\la{Kincomp}
\ee

\be     
H^i =\fr{1}{\sqrt{2}}
\left\{
K_{25}+i K_{30},
K_{26}+i K_{31},
K_{27}+i K_{32},
K_{28}+i K_{33},
K_{29}+i K_{34}
\right\}
\la{HLincomp}
\ee
To get the Realized form $W_R$, we just substitute (\ref{Kincomp}) and 
(\ref{HLincomp})   into (\ref{gensolsu5}) and (\ref{firstlineofVr}).
So our scalar potential is simply
   \be
V_{\rm Real\; After\;ET}
=
e^{\fr{1}{2} \k^2 \lt ( 
  \sum_{j=1}^{34}   K^i K^i
 \rt )}   
\lt \{
\fr{1}{2}  \sum_{j=1}^{34} \lt ( \fr{\pa  W_{\rm R} }{\pa K^i} \rt )^2
\rt \}
\la{complexV44left}\ee
and our superpotential is simply:
\be
W_R = 
e^{-\frac{1}{8}  \kappa ^2  \lt ( 
  \sum_{j=1}^{34}   K^i K^i
 \rt )}  M_P^{\fr{3}{2}}  2^{-\fr{3}{4}}
\sqrt{ g_1 \Tr( KKK) - g_2  H \cdot K \cdot  \oH      }   
\la{wusingallKhere}
\ee

\refstepcounter{orange}
{\bf \theorange}.\;
{\bf The Field Shift:}
\la{fieldshift} Now we  can look for shifts of the scalar fields that leave the scalar potential
(\ref{complexV44left}) at zero energy.  
First we perform the following shift in $W_R$ in (\ref{wusingallKhere}):
\be
\left\{K_{11}\to K_{11}+M_{\text{K1}},K_{12}\to K_{12}+M_{\text{K2}}
,K_{29}\to K_{29}+M_{\text{K9}}\right\}
\la{Shifts}\ee 
In the above, $M_{\text{K1}},M_{\text{K2}},M_{K9}$ are the three VEVs that we assume for these fields.
 The zero cosmological constant is maintained at this tree level by insisting that the VEV  of the expression V in (\ref{complexV44left})  is still zero after the shifts. This is familiar from gauge symmetry breaking in rigid or global SUSY, uncoupled to supergravity. 
  This requirment  is trivial for all variables except $K_{11}$,  $K_{12}$ and $ K_{29}  \equiv K_{H5}$.  This results in three equations.  
\be
\lt < \fr{\pa W_R}{\pa K_{12}}\rt >
=
\lt < \fr{\pa W_R}{\pa K_{11}}\rt >
= \lt <\fr{\pa W_R}{\pa K_{29}}\rt >
=0
\la{veviszero2}
\ee

\refstepcounter{orange}
{\bf \theorange}.\;
{\bf Introducing a Scale $r$:}
\la{scaling}
For the three equations in paragraph \ref{fieldshift}, we take 
\be
g_2 \ra r g_1
\la{lintrans}
\ee
 This quickly reduces those equations to a dependency on $r$ alone, provided that $g_1 \neq 0$.   
Then here are the equations from paragraph \ref{fieldshift} after removing $g_1,g_2$ with the transformation (\ref{lintrans}):

 \be
  \left\{3 r M_{\text{K9}}^2 \left(2 \sqrt{15} M_{\text{K1}} M_{\text{K2}}-10 M_{\text{K1}}^2+20
   M_P^2\right)-8 \sqrt{15} M_{\text{K1}} M_{\text{K2}} \left(9
   M_{\text{K1}}^2+M_{\text{K2}}^2-36 M_P^2\right)=0,
   \ebp
   4 \sqrt{15} \left(9 M_{\text{K1}}^2 \left(4
   M_P^2-2 M_{\text{K2}}^2\right)+12 M_{\text{K2}}^2 M_P^2-2 M_{\text{K2}}^4\right)-3 r
   M_{\text{K9}}^2 \left(10 M_{\text{K1}} M_{\text{K2}}+\sqrt{15} \left(4 M_P^2-2
   M_{\text{K2}}^2\right)\right)=0,
   \ebp
   3 r M_{\text{K9}} \left(\sqrt{15} M_{\text{K2}}-5
   M_{\text{K1}}\right) \left(M_{\text{K9}}^2-4 M_P^2\right)-4 \sqrt{15} M_{\text{K2}}
   M_{\text{K9}} \left(9 M_{\text{K1}}^2+M_{\text{K2}}^2\right)=0\right\}
   \ee

\refstepcounter{orange}
{\bf \theorange}.\;
{\bf Introducing a tiny parameter $f$:}
\la{tinyf}
Then the following\footnote{The three equations have 15 solutions \ci{mathnotebooks}.  Four of them lead to $
SU(5) \ra SU(4)$
and four more lead to $
SU(5) \ra SU(3) \times U(1)$. The other 7 have $M_{K9}\ra 0$, which is less interesting.  We choose one solution (Sol15) that leads to $
SU(5) \ra SU(3) \times U(1)$ and we assume that the other three that lead to $
SU(5) \ra SU(3) \times U(1)$ lead to similar results below.} is a solution of the equations in Paragraph \ref{scaling}:
\be
\left\{M_{\text{K1}}\to \frac{3 \sqrt{r} (r+2) M_P}{2 \sqrt{r^3-9 r^2+72 r+108}},M_{\text{K2}}\to
   \frac{\sqrt{15} (r-6) \sqrt{r} M_P}{2 \sqrt{r^3-9 r^2+72 r+108}},
\la{req23}   \ebp
   M_{\text{K9}}\to -3 \sqrt{2}
   \sqrt{-\frac{(r-18) (r+2)}{r^3-9 r^2+72 r+108}} M_P
   \right\}
  \la{req11}
   \ee
Now we define $f$, which is a very small ratio\footnote{See Paragraph \ref{renormgp} below.}:
\be
f= \fr{M_{K9}}{M_P} = \fr{M_{W}}{g_5 M_P} =  \fr{ 3.35 \times 10^{-17}}{  g_5} 
\la{valueoff}
\ee
Then the substitution (\ref{req11}) becomes
\be
-\frac{18 (r-18) (r+2)}{r^3-9 r^2+72 r+108}=f^2
 \la{eqforgsfromsol16andZafterr}
 \ee
The equation (\ref{eqforgsfromsol16andZafterr}) is a cubic equation for $r$ in terms of $f$. Mathematica can easily find the three solutions exactly in symbolic form, but those look complicated.  
They are easier to understand when we get 
\ma\  to expand them as power series
 in the tiny fraction $f$:  
 \be
 \left(
r \ra -\frac{18}{f^2}-7+\frac{110 f^2}{9} +O\left(f^4\right);\; r \ra 18-12 f^2+O\left(f^4\right) 
  ; \;   r \ra
    -2-\frac{2 f^2}{9} +O\left(f^4\right) 
\right)
\la{wect}
 \ee
\refstepcounter{orange}
{\bf \theorange}.\;
{\bf Three different results from the three solutions for the scale $r$:}
\la{uniquer}
So there are three different ways to obtain $M_{K9}$ in (\ref{req11}) here.  Any of the three solutions in
(\ref{wect})   yield $f^2$ in (\ref{eqforgsfromsol16andZafterr}) and so $ M_P f$ for  (\ref{req11}). But they will not give the same values for $M_{K1}$ and $M_{K2}$ in (\ref{req23}).  What they give, in the same order as in (\ref{wect}), is:
   \be
\left\{M_{\text{K1}}\to 
-\frac{3
   M_P}{2},
   M_{\text{K2}}\to 
    -\frac{\sqrt{15} M_P}{2},M_{\text{K9}}\to -f
   M_P\right\},
 \la{vev1}
   \eb
   \left\{M_{\text{K1}}\to 
    \frac{\sqrt{15}
   M_P}{2},M_{\text{K2}}\to \frac{3 M_P}{2},M_{\text{K9}}\to -f
   M_P\right\},
 \la{vev2}  \eb
   \left\{M_{\text{K1}}\to -\frac{f^2 M_P}{6 \sqrt{10}}, 
   M_{\text{K2}}\to 
   -\sqrt{6} M_P,M_{\text{K9}}\to -f M_P\right\}  
   \la{vev3}\ee
   
Note that the value of $M_{K1}$ is of order $M_{P}$  for the first two cases  (\ref{vev1}) and (\ref{vev2}).  These come from the first two choices  in (\ref{wect}).
However  the value of $M_{K1}$ is of order $f^2 M_{P}$
  for  the third case   (\ref{vev3}),  which comes from the third choice in (\ref{wect}).
   Now we will look at the Vector Boson Masses, and we will see that the only case that has a chance of meeting experiment is the third case   in (\ref{wect}), together with its consequence (\ref{vev3}).

\refstepcounter{orange}
{\bf \theorange}.\;
{\bf
Yang Mills Vector Fields:}
\la{vecbosons}
The VEVs above determine the masses of the vector bosons.  First we define: 
\be
V=\left(
\begin{array}{ccccc}
 V_3+\frac{V_8}{\sqrt{3}}-\frac{2 V_{12}}{\sqrt{15}} & V_1+i V_2 & V_4+i V_5 & V_{13}+i V_{14}
   & V_{19}+i V_{20} \\
 V_1-i V_2 & -V_3+\frac{V_8}{\sqrt{3}}-\frac{2 V_{12}}{\sqrt{15}} & V_6+i V_7 & V_{15}+i V_{16}
   & V_{21}+i V_{22} \\
 V_4-i V_5 & V_6-i V_7 & -\frac{2 V_8}{\sqrt{3}}-\frac{2 V_{12}}{\sqrt{15}} & V_{17}+i V_{18} &
   V_{23}+i V_{24} \\
 V_{13}-i V_{14} & V_{15}-i V_{16} & V_{17}-i V_{18} & V_{11}+\sqrt{\frac{3}{5}} V_{12} & V_9+i
   V_{10} \\
 V_{19}-i V_{20} & V_{21}-i V_{22} & V_{23}-i V_{24} & V_9-i V_{10} & \sqrt{\frac{3}{5}}
   V_{12}-V_{11} \\
\end{array}
\right)
\ee

\refstepcounter{orange}
{\bf \theorange}.\;
{\bf
Mass Term for Vector Bosons:}
\la{vecbosonmasses}
These arise from the following terms in the action \ci{ross,GUT}: 
\be
 \fr{1}{2}g_5^2 {\rm Tr} \lt (\lt [ V , S \rt ] \lt [ V , \oS \rt ] \rt ) 
\ra \lt <  \fr{1}{4}g_5^2 {\rm Tr} \lt (\lt [ V , K \rt ] \lt [ V , K \rt ] \rt ) \rt >
\eb
-  g_5^2  \lt ( \oH_L V V  H_L \rt ) 
\ra 
\lt <
- \fr{1}{2} g_5^2  \lt (  \oH V V  H \rt ) 
\rt >
;\; 
-   g_5^2  \lt (\oH_R V V  H_R \rt ) 
\ra 
\lt <
- \fr{1}{2}  g_5^2  \lt (H V V  \oH \rt )
\rt >\ee 
To find the eigenvalues of the Vector Bosons, and their multiplicities, we add the right hand sides of the above  together and we call the resulting expression $-\fr{1}{2}(V \cdot M_V^2 \cdot V)$.  
Then we implement the \ET s.  Next we substitute the expressions (\ref{Kincomp}) and 
(\ref{HLincomp}).  Then we perform the shift (\ref{Shifts}).  Then we set $K_i \ra 0$. Then we take two derivatives to generate the $24\times 24$ matrix $( M_V^2)_{ ab}$. The Mathematica  calculation results in the following eigenvalues of $( M^2)_{ab}$, with the following multiplicities, evaluated to the lowest  order of f: 
\be
r \ra -\frac{18}{f^2}
\Ra \left(
\begin{array}{cc}
 0 & 9 \\
64 g_5^2 M_P^2 & 6 \\
 \frac{8}{5} f^2 g_5^2 M_P^2 & 1 \\
36 g_5^2 M_P^2 & 2 \\
4 g_5^2 M_P^2 & 6 \\
\end{array}
\right); \; r \ra 18-12 f^2\Ra
\left(
\begin{array}{cc}
 0 & 9 \\
 60 g_5^2 M_P^2 & 6 \\
 \frac{8}{5} f^2 g_5^2 M_P^2 & 1 \\
60 g_5^2 M_P^2 & 2 \\
 f^2 g_5^2 M_P^2 & 6 \\
\end{array}
\right);
\la{six1}
\eb
 r \ra
    -2-\frac{2 f^2}{9}\Ra
\left(
\begin{array}{ccccc}
 0 & 40 g_5^2 M_P^2 & \frac{8}{5} f^2 g_5^2 M_P^2 & f^2 g_5^2 M_P^2 & 40 g_5^2 M_P^2 \\
 9 & 6 & 1 & 2 & 6 \\
\end{array}
\right)
\la{six2}
\ee

   The nine zero eigenvalues $0\to 9$ in (\ref{six1}) and (\ref{six2}), confirm that we have $SU(3) \times U(1)$ as our residual symmetry here. 
 The two solutions in (\ref{six1}), which come from  the first two solutions in (\ref{wect}), are unsatisfactory.  They both\footnote{The choice $ r \ra 18-12 f^2$ also leads, in the second item in (\ref{six1}),  to a low mass for one of the $X,Y$ complex triplet vector bosons.} yield totally unsatisfactory masses of order $M_P$ for the $W$ vector boson (the multiplicity 2 item). 
The only satisfactory solution is (\ref{six2}).  It comes from 
$ r \ra
    -2-\frac{2 f^2}{9}$, which is the third solution in (\ref{wect}).

\refstepcounter{orange}
{\bf \theorange}.\;
{\bf
Renormalization Group Issues and a Value for $g_5$:}
\la{renormgp}
Here are the squared masses of  the Z 
and  the W mass, from  (\ref{six2}): 
\be
 M_Z^2
=
\frac{8}{5} g_5^2  M_{P}^2 f^2
; \; M_W^2=
g_5^2
   M_P^2 f^2
\ee It follows that:
 \be
 \fr{ M_W}
{  M_{P}} =  f  g_5 =   3.35 \times 10^{-17} 
;\;  \cos^2 \q_{W}= \fr{M_W^2}{M_Z^2}=
\fr{5}{8}
\Ra
\sin^2 \q_{W}=
\fr{3}{8}
\la{valoffg5}
\ee
As is well known, the naive value of $\sin^2 \q_{W}$ in this SU(5) model does not agree with experiment. 
This feature has been much discussed in rigid versions of the SU(5) SUSY GUT and the renormalization group is clearly relevant here \ci{Weinberg3,Weinberg2}. 
Now we also have the problem of separating $g_5 f$ into the two components  $g_5$ and $f$.  This separation is needed in order to convert the results into specific mass predictions. We get different values for $g_5 f$,  depending on whether we try to fit $M_W$ or $M_Z$.  We cannot fit both. 
Here we choose to fit $M_W$.
In the Tables in Paragraph (\ref{conclusion}), we leave this question open. Our general assumption \ci{Weinberg3,Weinberg2} is that $g_5 \approx1$.  This calls for further analysis.

\refstepcounter{orange}
{\bf \theorange}.\;
{\bf Scalar Field Mass Matrix:}
\la{scalarfieldmassmatrix} 
Consider the real symmetric matrix:
\be
M_{ij} =  
\lt < {\fr{\pa^2 W_R }{\pa K_i \pa K_j}} \rt > 
;\;  {i = 1 \cdots 34},{j = 1 \cdots 34}
\la{halfhiggsmat2}\ee
This can be used to generate:
\be
{\rm M}_{\rm 
Scalar\;ij}^2=  \lt < {\fr{\pa^2 V_R }{\pa K_i \pa K_j}}  \rt > 
 =
 e^{ \fr{1}{2}\k^2 \lt ( 
M_{K9}^2+M_{K1}^2+M_{K2}^2
 \rt )}
M_{ik} M_{jk}
\la{higgsmasssqmat} \ee
where we use
(\ref{firstlineofVr})  or (\ref{complexV44left}) for $V_R$.
Note that for our chosen solution in (\ref{req23}) and  (\ref{req11}) 
    we have \ci{mathnotebooks} 
\be
\lt < e^{\k^2 \cK} \rt > \ra  e^{ \fr{1}{2}\k^2 \lt ( 
M_{K9}^2+M_{K1}^2+M_{K2}^2
 \rt )}=e^{3}
 \la{ethreehalf}
\ee
This mass squared matrix for the scalar bosons  (\ref{higgsmasssqmat}) 
 is necessarily positive semi-definite.
 It is easier to work out the  real symmetric matrix $M_{ij}$ in (\ref{halfhiggsmat2}) than it is to work out 
 (\ref{higgsmasssqmat}) directly.  We work out $M_{ij}$  and then let \ma\ find its eigenvalues.   . Then we simply square those eigenvalues and multiply by the factor
  $ e^{3}$, which comes from (\ref{ethreehalf}).
 This yields the same result as taking (\ref{complexV44left}), because of the equations 
 (\ref{veviszero2}).  
 
\refstepcounter{orange}
{\bf \theorange}.\;
{\bf Eigenvalues of the Scalar Field Mass Matrix:}
\la{eigsofscalarfieldmassmatrix} 
  There are six non zero real eigenvalues for  ${\rm M}_{\rm 
Scalar\;ij}^2$. 
Firstly, there are 15 zero eigenvalues.  Then there are masses for a real octet, a complex triplet and a complex singlet.
Then there are 3 real Higgs boson masses which are the three cube roots of a complicated cubic equation. We can reduce this complicated set of masses to simple ones however.
First we substitute  $ g_2 \ra r g_1$ in the expressions for the Masses in $M_{ij}$.  Then we substitute 
$r\to -2-\frac{2 f^2}{9}-\frac{29 f^4}{405}-\frac{887 f^6}{36450}+O\left(f^{8}\right)
$ in accord with the results in Paragraph \ref{vecbosonmasses}. The remaining dependence on $g_1$ turns out to be exactly linear for the square of  $M_{ij}$, as seen below in Equation (\ref{masandmultscalars}).  
We find \ci{mathnotebooks} that the squared masses and their multiplicities reduce simply to the following, to lowest order in $f$:
\be
{\rm M}_{\rm 
Scalar}^2\ra \left(
\begin{array}{ccccccc}
 0 & \frac{15}{8} \sqrt{5} M_{\text{SP}}^2 & \frac{5}{24} \sqrt{5} M_{\text{SP}}^2 &
   \frac{15}{8} \sqrt{5} M_{\text{SP}}^2 & \frac{3 M_{\text{SP}}^2}{2 \sqrt{5}} & \frac{15}{8}
   \sqrt{5} M_{\text{SP}}^2 & \frac{f^4 M_{\text{SP}}^2}{54 \sqrt{5}} = M_H^2\\
 15 & 8 & 6 & 2 & 1 & 1 & 1 \\
\end{array}
\right)
\la{masandmultscalars}
\ee
In the above we use an abbreviation $M_{\text{SP}}=e^{3/4} \sqrt{-g_1 M_P^2}$ for the `SuperPlanck' mass, which we define below in (\ref{spmass}).
The sum is 15 Goldstone+  8 {\rm Octet} + 6 Complex Triplet+ 2 Doublet  +  3 {\rm Higgs}   = 34 real Higgs Bosons.
Since we need the known Higgs Mass to be the lightest of the three neutral masses, it follows that, as shown in the last term in
(\ref{masandmultscalars}):
\be
M_{\rm Higgs}^2 \equiv h^2 M_P^2=\frac{f^4 M_{\text{SP}}^2}{54 \sqrt{5}} =
\frac{f^4 }{54 \sqrt{5}} \lt ( e^{3/4} \sqrt{-g_1 M_P^2}\rt )^2=
- \frac{f^4 }{54 \sqrt{5}} e^{3/2}  g_1 M_P^2
\la{forH}
\ee
Our unit of mass here is the SuperPlanck mass\footnote{This closely resembles the neutrino see saw mechanism \ci{neutrinoseesaw}. The Higgs mass is tiny compared to the Planck mass, just as the neutrino mass is tiny compared to the Electroweak masses.}.  It takes the value
\be
M_{\text{SP}}=e^{3/4} \sqrt{-g_1 M_P^2}= 
 1.22\times 10^{36} g_5^2\; {\rm GeV}
=
2.18 \times 10^{12} g_5^2 \;{\rm grams}
= 2.18  \; g_5^2  \;{\rm Megatonnes}
\la{spmass}
\ee
where we determined the value of $g_1$, using (\ref{valoffg5}) and (\ref{forH}), to be:
 \be
g_1=-e^{-3/2}  \frac{54 \sqrt{5} M_{\rm Higgs}^2} {f^4   M_P^2} = 
 - e^{-3/2}  \frac{54 \sqrt{5} g_5^4   } {(g_5 f)^4  } h^2
  = -5.81 \times 10^{34} g_5^4
\ee

\refstepcounter{orange}
{\bf \theorange}.\;
{\bf The Gravitino Mass Term:}  
\la{calofgravitinomass}
The Gravitino mass term can be found in  \ci{freepro}: 
\be
{\cL}^{(2)}
= \fr{1}{2}  \k^2 e^{   \k^2 \fr{1}{2} \cK  }
W \ov \y_{\m R} \s^{\m \n}\y_{\n R}
\ee
So the mass squared is, from (\ref{gensolsu5})
\be
M^2_{\rm Gravitino}= \lt <   \k^4 e^{   \k^2  \cK  } W_R^2
\rt > = \lt <
  \k^4 
e^3 M_P^{3}
\lt |  g_1 \Tr( KKK) - g_2 H \cdot K \cdot  \oH         
\rt |
\rt >\ee
Now we put in the  VEVs, depending on  $g_5$, as noted in Paragraph \ref{renormgp}, to get \ci{mathnotebooks}:
\be
M_{\rm Gravitino}  =
2 \times 10^{36} g_5^2
\; {\rm GeV} \approx 3.57 \;g_5^2\; {\rm Megatonnes}
\ee

\refstepcounter{orange}
\la{mastereqpara}
{\bf \theorange}.\;
{\bf The Master Equation:}  
The origin of  `Suppressed SUSY' is in the Master Equation for SUSY. 
As in any gauged quantum field theory, the Master equation generates all the Ward identities of the theory that arise from its symmetries.  This Master equation takes the form:
\be
\cM\lt [ \cA\rt ] \equiv 
\int d^4 x \lt \{
\fr{\d \cA}{\d B^I}
\fr{\d \cA}{\d \S_I }
+
\fr{\d \cA}{\d {\cal F}^{I} }
\fr{\d \cA}{\d  S_{I} } 
\rt \}=0
\la{masterequat}
\ee

Here $B^I$ are  (Grassmann even) `Bosons' and $ {\cal F}^{I}$ are  (Grassmann odd) `Fermions'.  Then $\S_I$ are Grassmann odd Zinn sources\footnote{`Zinn sources' have sometimes been called `Antifields', but they are sources, not fields, and they are not anti anything, so we call them Zinn sources instead. These  play a crucial role in Suppressed SUSY, and their source character is crucial.} and $  S_{I}  $ are Grassmann even Zinn sources. One must include all the fields here, and $I$ is a collective sort of index.  The Physical Bosons  are   $z^{\a},A^A_{\m}, e^a_{\m}$ and the Physical Fermions are 
$\c, \lam,\y_{\m}$.  Many kinds of ghosts and antighosts and gauge fixing terms are also present, and auxiliaries are integrated.

\refstepcounter{orange}
{\bf \theorange}.\;
{\bf The Master Equation requires Supergravity:}  
The easiest way to see this is to recall that one needs to integrate over all fields to get the Master Equation, and it is essential that the starting transformations be nilpotent.  Ghosts and antighosts are fields for this purpose, as is well known from the Yang Mills theory. Any SUSY theory requires closure using the fact that two susy transformations yield a translation. For rigid SUSY, this means that the translation and SUSY ghosts are space-time constants.  But an attempt to keep the supergravity ghosts constant means that one cannot integrate them as fields. So they need to be spacetime dependent to formulate the Master Equation.  But that means that  \SG\ is present.

\refstepcounter{orange}
{\bf \theorange}.\;
{\bf The Nilpotent Operator:}  
Becchi, Rouet and Stora, with help from Zinn-Justin and Tyutin, and many others, noted that there is a fundamental feature here \ci{becchiarticles1,becchiarticles2}.  The form above gives rise to a nilpotent operator as follows:

\be
\d \equiv 
\int d^4 x \lt \{
\fr{\d \cA}{\d B^I}
\fr{\d  }{\d \S_I }
+
\fr{\d \cA}{\d \S_I }
\fr{\d  }{\d B^I}
+
\fr{\d \cA}{\d {\cal F}^{I} }
\fr{\d }{\d  S_{I} } 
+
\fr{\d \cA}{\d  S_{I} } 
\fr{\d }{\d {\cal F}^{I} }
\rt \}
\ee
Because (\ref{masterequat}) yields zero, it follows that:
\be
\d^2=0
\ee

\refstepcounter{orange}
{\bf \theorange}.\;
{\bf Generating Functionals and Counterterms are Cocycles of $\d$:}  
It is easy to show that the one-loop contribution to the 1PI Generating Functional ${\cal G}^1$, and the local one-loop contribution ${\cal A}^{1}$ to the renormalized 
action should satisfy:
\be
\d 
{\cal G}^1
=
0
;\; \d 
{\cal A}^{1}
=
0
\ee
The solution of the second equation above has the form
\be 
{\cal A}^{1}
={\cal H}^{1}
+\d {\cal B}^{1}
\ee
where ${\cal H}^{1}$ are local invariants in the cohomology space of $\d$,  and the local `coboundaries' $ {\cal B}^{1}$ are basically terms that yield field and source renormalizations.  One can then iterate this situation to any order using results like those in \ci{dixonnucphys}.  Modulo interesting issues like anomalies, the result of this is that the action that renormalizes the theory to all orders has a simple relation with the original action.  It is generally the same as regards the field and Zinn source content and structure, but there is room for renormalization of the coefficients. For the non-renormalizable theory here, there are new invariants too.

\refstepcounter{orange}
{\bf \theorange}.\;
\la{genfunc}
{\bf The Generating Functional for the Higgs Sector:}  
For the Higgs sector, the old field variables and the old Zinn sources are:
$
{\rm Fields}:\; H_L^i,H_{Ri}, S^a;\; {\rm Zinns}: \;
 \G_{L}^i,
  \G_{Ri},
  \G_S^a; \; (i = 1\cdots 5, a= 1\cdots 24)
$. These are all complex. They appear in the old action in 
$
\cA_{\rm Old\;Zinn} =\int d^4 x \lt \{
\G_{L}^i
\d H_{Ri}
+  \G_{Ri}
\d H_L^i 
+ \G_S^a
\d S^a 
\rt \} + *
+ \cdots $
where the missing terms include non-linear functions of the sources resulting from integration of 
auxiliaries.

The new Field variables, Zinns 
and antighosts are:
${\rm Fields}:\; H^i, K^a; \; {\rm Zinns}: \;
\G_i, \G^a
,J^i, J^a
;\; {\rm Antighosts}: \;
\h_i, \h^a; \; (i = 1\cdots 5, a= 1\cdots 24)
$.  $H^i,\G_i,J^i,\h_i$ are complex. $K^a,\G_a,J^a,\h^a$ are real. These appear in the new action in $
\cA_{\rm New\; Zinn} =\int d^4 x \lt \{
 \G_{i}
\d H^i 
+ J_{i}
\d \h^i +*
+ \G^a
\d K^a 
+ J^a
\d \h^a 
+ \cdots 
\rt \} 
$
where the missing terms include non-linear functions of the Sources resulting from integration of 
auxiliaries.

Here is the Generating Functional   \ci{four,five,goldstein, LandauLifmechanics} for the Higgs Sector that we will use here.  It is a function of the old fields and the new Zinns (except J-type sources) and antighosts:
\be
\cG_{\rm Higgs}=
 \int d^4 x \lt \{
\fr{1}{\sqrt{2}}
\lt ( 
H_L^i
+\ov H_R^i \rt)
\G_i
+
\fr{i}{\sqrt{2}}
\lt ( 
H_L^i
-\ov H_R^i \rt)
\h_i
+  S^a 
\fr{1}{\sqrt{2}}\lt(
 \G^a
 +  i \h^a
\rt )
\rt \} + *
\ee
The new fields and the new $J$-type sources are obtained as functions of the old fields as follows:
\be
H^i= \fr{\d \cG}{\d \G_i}=
\fr{1}{\sqrt{2}}
 \lt ( 
H_L^i
+\ov H_R^i \rt)
;
\;J^i= \fr{\d \cG}{\d \h_i}=
\fr{i}{\sqrt{2}}
 \lt ( 
H_L^i
-\ov H_R^i \rt)
;\;
\eb
 K^a 
= \fr{\d \cG}{\d    \G^a}=
 \fr{1}{\sqrt{2}}\lt(
   S^a
+  \ov S^a 
\rt )
;\;
J^a 
= \fr{\d \cG}{\d    \h^a}=
 \fr{i}{\sqrt{2}}\lt(
   S^a
-  \ov S^a 
\rt )
 \ee
We can combine these to write the old fields in terms of the new fields and  the new $J$-type sources:
\be
H_L^i=
\fr{1}{\sqrt{2}}
\lt (H^i -i J^i\rt )
;\;\ov H_R^i 
=\fr{1}{\sqrt{2}}
\lt (H^i +i  J^i\rt )
;\;S^a 
=
 \fr{1}{\sqrt{2}}\lt(
K^a 
-i
J^a
\rt )
;\;\oS^a
=
 \fr{1}{\sqrt{2}}\lt(
K^a 
+i
J^a
\rt )
 \ee
   Note that for the construction of the quantized action, the two complex fields $H_L, H_R$ are replaced by one complex field $H$, and the complex Field $S$ in the adjoint representation of the SU(5) gauge theory, is  replaced by a real field  $K$.
The old Zinn sources are obtained from:
\be
\G_{Ri}= \fr{\d \cG}{\d H_L^i}=
\fr{1}{\sqrt{2}}
 \lt ( 
\G_i + i \h_i 
\rt)
;\;
\G_{L}^{i}= \fr{\d \cG}{\d H_{Ri}}=
\fr{1}{\sqrt{2}}
 \lt ( 
\ov \G^i - i \ov \h^i 
\rt)
;\; \G^a_{S } 
= \fr{\d \cG}{\d   S^a}=
 \fr{1}{\sqrt{2}}\lt(
 \G^a 
 +  i \h^a 
\rt )
\ee
The latter are relevant to the cohomology etc. but we do not need them here for the present Lagrangian construction. 

\refstepcounter{orange}
{\bf \theorange}.\;
\la{genfunc2}
{\bf The Generating Functional for the Gauginos, and the Meaning of the Master Equation:} 
 The Generating Functionals for the other scalar and spin $\fr{1}{2}$ sectors are simpler, at least for now. For example if the old Gauginos and Zinn sources are $\lam^a, Y^a$ and the new Zinn Sources and antighosts are $\Lam^a,M^a$
we could take: $
\cG_{\rm Gauginos}=
 \int d^4 x \lt \{
\lam^a
  M^a 
\rt \}  
;\;  \Lam^a 
= \fr{\d \cG}{\d   M^a}=
\lam^a
;\;  Y^a 
= \fr{\d \cG}{\d
\lam^a}= M^a
$,
where all the sources and fields are Majorana spinors ($Y^a,M^a$ are Grassmann even and $\lam^a,\Lam^a$ are Grassmann odd).
Note in particular that: 
 \be
 \int d^4 x Y^a
\d \lam^a   
=  \int d^4 x Y^a
 \pa_{\m} V^a_{\n}  \g^{\m\n}C + \cdots\ra 
 \int d^4 x M^a
 \pa_{\m} V^a_{\n}  \g^{\m\n}C  + \cdots
 \ee
  The last term $ \int d^4 x M^a
 \pa_{\m} V^a_{\n}  \g^{\m\n}C  $ does not  do very much.  There is no kinetic term here for the anti ghost  $M^a$, so it does not propagate.   Gauge fields are not so simple.
Here is a useful remark \ci{becchiremark}:
\begin{verse}
The Master Equation is the most general quantum mechanical solution for the problem of unphysical degrees of freedom, whenever they appear.
\end{verse}
In this paper, we take this statement farther, to conjecture  that those \ET s of the Master Equation which  change the functionality of scalar bosons and spin $\fr{1}{2}$ fermions, also satisfy the quantum mechanical requirements.
Note that the full \ET s are linear, homogeneous and invertible, so that the full Action, with Zinn sources, simply gets various terms renamed. 
  These formal aspects of Suppressed SUSY clearly need careful attention.

\refstepcounter{orange}
{\bf \theorange}.\;
{\bf Conclusion:}
\la{conclusion}
Suppressed SUSY is a very simple idea, but it has complicated consequences.
Here is a summary.  In accord with Paragraph \ref{renormgp}, $g_5 \approx 1$ is left undetermined:
\be
\begin{array}{|c|c|}
 \hline
\multicolumn{2}{|c|}{\rm  Parameters }\\
 \hline
  f g_5  =  M_W/M_P  &h = M_H/M_P  \\
 \hline
     3.35 \times 10^{-17}  g_5 &5.2 \times 10^{-17}   \\
\hline
\end{array}
\hspace{.1cm}
\begin{array}{|c|c|}
 \hline
\multicolumn{2}{|c|}{\rm Scales }\\
 \hline
 g_1 &r  = {g_2}/{g_1}   \\
 \hline
 -5.81 \times 10^{34} g_5^4
  &   -2- 2 f^2/9  \\
\hline
\end{array}
\hspace{.1cm}
\begin{array}{|c|c|}
\hline
\multicolumn{2}{|c|}{\rm Mass\; Names\; (GeV) }\\
 \hline
  M_P & M_{\rm SP} \\
 \hline
2.4 \times 10^{18} &  1.22\times 10^{36} g_5^2\ \\
\hline
\end{array}
\ee
\be
 \begin{array}{|c|c|c| }
\hline
\multicolumn{3}{|c|}{\rm VEVs\; (GeV)} \\
 \hline
M_{\text{K9}} & M_{\text{K1}}  &M_{\text{K2}}    \\
 \hline
 -f M_{P}& -\fr{M_P f^2}{6
   \sqrt{10}}& -\sqrt{6}
   M_P
    \\
\hline
\end{array}
\hspace{.1cm}
 \begin{array}{|c|c|c| }
\hline
\multicolumn{3}{|c|}{\rm Electroweak\; Masses\; (GeV)} \\
 \hline
{\rm M}_Z  = \fr{8 }{5}f g_5  M_P &{\rm M}_W =f g_5 M_P&{\rm M}_H  =h M_P   \\
\hline
\end{array}
\hspace{.1cm}
\begin{array}{|c|c|}
\hline
\end{array}
\ee

\be
\begin{array}{|c|c|}
\hline
\multicolumn{2}{|c|}{ \rm Planck\; Masses\; (GeV) }\\
 \hline
{\rm X}   &{\rm Y}     \\
\hline
 2\sqrt{10}  g_5 M_P &
2 \sqrt{10}  g_5 M_P \\
\hline
\end{array}
\hspace{.1cm}
\begin{array}{|c|c|c|c|c|c|}
\hline
\multicolumn{6}{|c|}{\rm Super\; Planck\; Masses \; (GeV)}\\
 \hline
{\rm H_{\rm Octet}} &{\rm H_{Triplet} }&{\rm H^+}&{\rm Higgs}_2  &{\rm Higgs}_3   
&{\rm Gravitino}\\
\hline
2.05  M_{\rm SP} &0.68   M_{\rm SP}&   2.05 M_{\rm SP}&  .81 M_{\rm SP} &2.05   M_{\rm SP} &  1.64 M_{\rm SP} 
\\
\hline
\end{array}
\la{sumtable}
\ee

After the \ET s, the Action, written in terms of the remaining quantized fields alone,  looks smaller and simpler.  
There is a natural way to get a zero \cc\ (at tree level), as was demonstrated using the `square root'
mass style superpotential above in (\ref{fresefas}), which generated the scalar potential (\ref{firstlineofVr}), while  (\ref{complexV44}) is identically zero. This depends on $g_1,g_2$ and $M_P$.
 After the \ET\ here, this model has many possible vacuum states, all with zero energy and zero \cc.  We chose to examine one where SU(5) breaks down to \bsg.
There is a natural see-saw which creates three levels of mass, and all the masses in the Higgs/ Gauge sector are determined from the $g_1,g_2$ in (\ref{gensolsu5}),   $M_P= \fr{1}{\k}$ and $g_5$.
 The VEV equations here that generate the Vector Boson mass matrix are functions only of the ratio $r=\fr{g_2}{g_1}$ for this model. 
 
  As shown in Paragraphs \ref{uniquer} and \ref{vecbosonmasses}, there are three possible solutions of the VEV equations that arise for the theory, but only one of them has a W mass comparable to the Z mass. This one requires that 
$r\to -2-\frac{2 f^2}{9}+O\left(f^{4}\right)$, where $g_5 f=\fr{M_W}{M_P}\approx 10^{-17}$, and then the masses of the X and Y vector bosons are of order $M_P$.
The Scalar mass squared matrix is a linear function of $g_1$, multiplied by a complicated function of $r$.  Inserting $r\to -2-\frac{2 f^2}{9}+O\left(f^{4}\right)$ yields a mass matrix for the six scalar boson masses. As shown in Paragraph \ref{eigsofscalarfieldmassmatrix}, setting the lowest mass neutral scalar to have the mass of the Higgs determines the value $g_1$ in terms of $f$ and $g_5$, and it then follows that the five other scalar boson multiplets, and the Gravitino (Paragraph \ref{calofgravitinomass}), all have masses of the SuperPlanck order
$M_{\rm SP} 
 \approx  1 {\rm Megatonne}
$.
It is simple to remove or add scalar or spinor fields, to vary and test the model.  Here we removed  half the scalar Higgs, and all of the Squarks,  Sleptons, Gauginos and Higgsinos.  We need to add some of them back to deal with the remaining issue of the weak angle etc.
With \sS, spontaneous breaking of SUSY is not necessary because superpartners can be removed, and so the need for the SSM \ci{haber}, with its invisible sector and explicit soft SUSY breaking   is gone.
 Once some Suppressed SUSY is present, one could expect mass splitting everywhere from loop effects. 
The masses found here seem encouraging for the ideas of Suppressed SUSY.  A priori, it might have been impossible to get the masses of the W, Z and Higgs sufficiently smaller than those of the other particles, given that there are very few parameters here.  
Nucleon decay has not been found at the level indicated by the SU(5)
 GUT as predicted by the   older methods  \ci{GUT}. However  the relevant masses (X, Y, Heavy Higgs) are predicted to be heavier using \sS, so this question is re--opened here.
Another feature that is quite strange here, and encouraging, is that it is possible to fit the Higgs mass with the see saw mechanism in Paragraph \ref{eigsofscalarfieldmassmatrix}. This arises naturally from the form of the superpotential.  The relevant matrix is very complicated, and the reason that this happens is not clear to the author. 
It is also possible of course to take more complicated versions of the superpotential as noted in footnote \ref{ftnt1} above.  What changes arise from that?

The Gravitino in this model is extremely massive ($\approx$ 1 Megatonne), and stable too. An improved  model would need to re-insert some fields to deal with the the weak angle, and also to keep the cosmological constant reasonable at loop order, if possible. That would probably create a decay mechanism for the Gravitino, maybe giving rise to a lighter LSP candidate for dark matter \ci{darkmatter}. At any rate, this theory needs examination at one loop, and perhaps, ultimately, an examination from the superstring point of view.

\begin{center}
 {\bf Acknowledgments}
\end{center}
\vspace{.1cm}

  I thank Doug Baxter, Carlo Becchi, Philip Candelas,  James Dodd, Mike Duff, Richard Golding,  Dylan Harries, Chris Hull, Sergei Kuzenko,  Pierre Ramond, Graham Ross, Peter Scharbach,    Kelly Stelle,  Xerxes Tata,  J.C. Taylor, Steven Weinberg,  and Peter West for stimulating correspondence and conversations.

\tiny
\articlenumber\\
\today
\end{document}

%% file: Johnmacros22.tex
\newcommand{\GUST}{{\rm SU(5) Grand Unified Supergravity Theory}}

 \newcommand{\ET}{Exchange Transformation}

\newcommand{\sS}{Suppressed SUSY}

\newcommand{\cM}{{\cal M}}

\newcommand{\ma}{{Mathematica}}
\newcommand{\SG}{{Supergravity}}

\newcommand{\cc}{{cosmological constant}}

\newcommand{\cG}{{\cal G}}
\newcommand{\cK}{{\cal K}}

\newcommand{\cA}{{\cal A}}

\newcommand{\PB}{Master Equation}

\newcommand{\bsg}{$SU(3)  \times U(1)$}

\newcommand{\cL}{\cal L}
\newcommand{\bt}{\begin{tabular}{c}}
\newcommand{\et}{\end{tabular}}

\newcommand{\eb}{\ee\be } 
\newcommand{\ebp}{\rt.\ee\be\lt.} 
\newcommand{\bmat}{\lt ( \begin{array} }
\newcommand{\emat}{  \end{array} \rt )}

\newcommand{\oH}{{\ov H}}

\newcommand{\oS}{{\ov S}}

\newcommand{\ED}{\end{document}}

\renewcommand{\a}{\alpha}	

\newcommand{\g}{\gamma}
\renewcommand{\d}{\delta}

\newcommand{\ve}{\varepsilon}

\newcommand{\h}{\eta}
\newcommand{\q}{\theta}
\renewcommand{\k}{\kappa}
\newcommand{\lam}{\lambda}
\newcommand{\m}{\mu}
\newcommand{\n}{\nu}

\newcommand{\s}{\sigma}

\renewcommand{\c}{\chi}
\newcommand{\y}{\psi}

\newcommand{\G}{\Gamma}

\newcommand{\Lam}{\Lambda}

\renewcommand{\S}{\Sigma}

\newcommand{\la}{\label}
\newcommand{\ci}{\cite}

\newcommand{\ds}{\documentstyle}	
\newcommand{\fr}{\frac}

\newcommand{\pa}{\partial}
\newcommand{\ov}{\overline}
\newcommand{\be}{\begin{equation}}
\newcommand{\ee}{\end{equation}}
\newcommand{\ba}{\begin{array}} 
\newcommand{\ea}{\end{array}}
\newcommand{\bea}{\begin{eqnarray}}
\newcommand{\eea}{\end{eqnarray}}
\newcommand{\ra}{\rightarrow}
\newcommand{\Ra}{\Rightarrow}

\newcommand{\lt}{\left}
\newcommand{\rt}{\right}

\newcommand{\Tr}{{\rm	Tr}	}

\newcommand{\ben}{\begin{enumerate}}
\newcommand{\een}{\end{enumerate}}
\newcommand{\bitem}{\begin{itemize}}
\newcommand{\eitem}{\end{itemize}}

%% file: 5221RS.bbl
\begin{thebibliography}{99}
\bibitem{nonrermorm} \sS\ does not, in general, respect the non-renormalization theorems, which are discussed in all of the SUSY literature. These go along with the features in Paragraph \ref{obvious}. It has often been argued that the non-renormalization theorems help to explain the `hierarchy problem', and that this supports the SSM.  This is an issue  that needs careful attention in the future, supposing that \sS\ yields some interesting results.
\end{thebibliography}
